\documentclass[useAMS,usenatbib]{mn2e}
\usepackage{natbib}
\usepackage{epsfig}
\usepackage{amsbsy}
\usepackage{hyperref}

\providecommand{\adsurl}[1]{\href{#1}{ADS}}

\newcommand{\mean}{\overline}
\newcommand{\lya}{Lyman-$\alpha$}

\title[Neutrino Signatures on the High Transmission Regions of the
Lyman-$\alpha$ Forest]
{Neutrino Signatures on the High Transmission Regions of the
Lyman-$\alpha$ Forest}

\author[F. Villaescusa-Navarro et al.] 
{F.~Villaescusa-Navarro$^{1,2,3}$, M.~Vogelsberger$^{2}$, M.~Viel$^{3,4}$, A.~Loeb$^{2}$
\\
$^1$ IFIC, Universidad de Valencia-CSIC, E-46071, Valencia, Spain (villaescusa@oats.inaf.it)\\
$^2$ Harvard-Smithsonian Center for Astrophysics, 60 Garden Street, Cambridge, MA, 02138, USA\\
$^3$ INAF-Osservatorio Astronomico di Trieste, Via G.B. Tiepolo 11, I-34131 Trieste, Italy\\
$^4$ INFN sez. Trieste, Via Valerio 2, 34127 Trieste, Italy \\}

\begin{document}
\maketitle
\begin{abstract}
We quantify the impact of massive neutrinos on the statistics of low
density regions in the intergalactic medium (IGM) as probed by the
\lya~forest at redshifts $z=2.2$--$4$. Based on mock but realistic
quasar (QSO) spectra extracted from hydrodynamic simulations with cold
dark matter, baryons and neutrinos, we find that the probability
distribution of weak \lya~absorption features, as sampled by \lya~flux
regions at high transmissivity, is strongly affected by the presence
of massive neutrinos. We show that systematic errors affecting the
Lyman-$\alpha$ forest reduce but do not erase the neutrino
signal. Using the Fisher matrix formalism, we conclude that the sum of
the neutrino masses can be measured, using the method proposed in this
paper, with a precision smaller than 0.4 eV using a catalog of 200
high resolution (S/N$\sim$100) QSO
spectra. This number reduces to 0.27 eV by making use of reasonable
priors in the other parameters that also affect the statistics of the
high transitivity regions of the Lyman-$\alpha$ forest. The
constraints obtained with this method can be combined with independent
bounds from the CMB, large scale structures and measurements of the
matter power spectrum from the Lyman-$\alpha$ forest to produce
tighter upper limits on the sum of the masses of the neutrinos.
\end{abstract}


\section{Introduction}  

Neutrino oscillation experiments revealed that neutrinos are not
massless particles. Since then a major effort has been dedicated to
measure or constrain neutrino masses. Current laboratory bounds
constrain the electron neutrino mass to $m_{\nu_e}<2.05$ eV
\citep{Lobashev,Kraus}. Cosmological bounds for the sum of all
neutrino masses are still significantly stronger: constraints from
WMAP7 alone yield $\Sigma_im_{\nu_i}<1.3$ eV \citep{Komatsu}, while
combined with large scale structure (LSS) measurements they constraint
the mass to $\Sigma_im_{\nu_i}<0.3$ eV \citep{Wang,Thomas,Concha,Reid,
  dePutter,Xia} . The tightest $2\sigma$ upper limit of
$\Sigma_im_{\nu_i}<0.17$ eV, is obtained by combining cosmic microwave
background (CMB) results, LSS and \lya~forest\citep{Seljak} data sets
(see \citep{Abazajian} for a summary of current and future neutrino
mass constrains). Among all the different observables the \lya$ $
forest is particularly constraining since it probes structures over a
wide range of redshift, in a mildly non-linear regime and at small
scales where the neutrino signature is present
\citep{Pastor,Rauch,Meiksin07}.

The dynamics of cosmological neutrinos is very different from that
of the dominant cold dark matter (CDM) component. The large velocity
dispersion of neutrinos suppresses their power spectrum of density
fluctuations at small scales, making the shape of the total power
spectrum a potential probe of neutrino masses. 

Previous studies have addressed the role of neutrinos in dark matter
halos \citep{Ma,Wong,Paco}, LSS \citep{Ma-Bert,Brandbyge_particle,
  Brandbyge_grid, Brandbyge_hybrid,Hannestad2,Viel, Bird} and the
intergalactic medium (IGM) \citep{Springel}, using both linear theory
and N-body/hydrodynamic techniques for the non-linear regime.  It has
been shown that on scales of 1-10 $h^{-1}$Mpc the non-linear
suppression is redshift, scale and mass dependent in a way that is
different from a naive extrapolation of linear theory.

\begin{figure}
\centering
\includegraphics[width=0.49\textwidth]{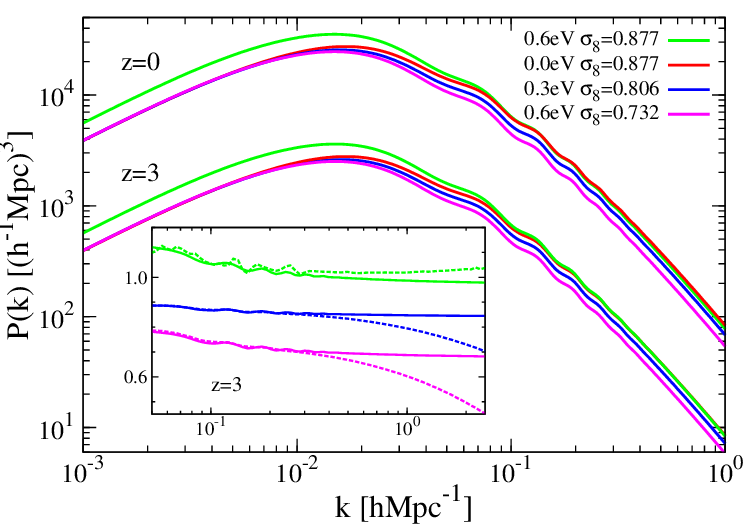}
\caption{Linear matter power spectra for different neutrino masses at $z=0$
  (upper lines) and $z=3$ (bottom lines). The inner panel shows the
  linear (solid lines) and non-linear (dotted lines) matter power spectrum at
  $z=3$ normalized to the case without neutrinos. On scales $k<0.03~h$
  ${\rm Mpc^{-1}}$ changes in the non-linear power spectra are driven
  by the differences in the linear power spectra.}
\label{power_spectrum}
\end{figure}

In this paper we study the effect of massive neutrinos on the
properties of low density regions or {\it voids} in the intergalactic
medium (IGM). Neutrinos have only a mild effect on dark
matter halos \citep{Ma,Wong,Paco}, since their large velocity
dispersion prevents their clustering on small scales. In contrast, we
find that the impact of neutrinos on void properties is much
stronger. Voids are relatively empty regions with
$\delta=\rho_m/\bar{\rho}_m-1$ ranging from almost $-1$ in their cores
to $\sim-0.7$ at radii $10-20$ Mpc at $z=0$ \citep{Void_comparison}.
By solving the dynamical equations for an isolated spherical top-hat
underdense perturbation, we find that neutrinos modify the evolution
of underdense regions by making them smaller and denser. Neutrinos
contribute to the interior mass of the underdense region delaying the
rate at which CDM is being evacuated from its interior and slowing
down the velocity of the shell surrounding it. We find that the
linearly extrapolated density contrast when the underdense region
enters into its non-linear phase decreases by $\sim 10\%$ for
neutrinos with $\Sigma_im_{\nu_i}\sim 1$ eV. Using the analytic model
presented in \citep{Sheth} we find that the statistics of voids depend
on both $\sigma_8$ and $\Sigma_i m_{\nu_i}$.  \lya~voids and their
dependence on other cosmological parameters have been investigated in
\citep{Viel_voids}. Here we focus on the dependence of void properties
on the sum of the neutrino masses.  We consider the \lya~signature of
low density regions, and introduce a new and simple statistical tool
(note that a similar observable was already studied in
\citep{Jordi,Fan}) that samples most of the IGM volume and appears to be
highly sensitive to neutrino masses.

\section{Numerical Method} 

\begin{table}
\begin{center}
\begin{tabular}{|c|c|c|c|c|c|c|}
\hline
Name & $\Sigma_i m_{\nu_i}$ (eV) & $\sigma_8$ ($z=0$) & $N_\mathrm{CDM}^{1/3}$ & $N_\mathrm{b}^{1/3}$ & $N_\nu^{1/3}$ \\  \hline  \hline 
S0 & $0.0$ & $0.877$ & $512$ & $512$ & $0$ \\  \hline
S0+ & $0.0$ & $0.928$ & $512$ & $512$ & $0$ \\  \hline
S0- & $0.0$ & $0.828$ & $512$ & $512$ & $0$ \\  \hline
S3 & $0.3$ & $0.948$ & $512$ & $512$ & $0$ \\  \hline
S3+ & $0.3$ & $0.877$ & $512$ & $512$ & $0$ \\  \hline
S3- & $0.3$ & $0.807$ & $512$ & $512$ & $0$ \\  \hline
S5 & $0.5$ & $0.755$ & $512$ & $512$ & $0$ \\  \hline
S6 & $0.6$ & $0.732$ & $512$ & $512$ & $0$ \\  \hline
S6+ & $0.6$ & $0.877$ & $512$ & $512$ & $0$ \\  \hline
S7 & $0.7$ & $0.709$ & $512$ & $512$ & $0$ \\  \hline
LR0 & $0.0$ & $0.877$ & $448$ & $448$ & $0$ \\  \hline
LR6 & $0.6$ & $0.732$ & $448$ & $448$ & $0$ \\  \hline
P0 & $0.0$ & $0.877$ & $512$ & $512$ & $512$ \\  \hline
P6 & $0.6$ & $0.732$ & $512$ & $512$ & $512$ \\  \hline
\end{tabular}
\end{center}
\caption{Summary of the N-body/hydrodynamic parameters of the
  simulations. The cosmological parameters are the same for all
  simulations and are given in the text.  $\Omega_{\rm M}=\Omega_{\rm
    cdm}+\Omega_{\rm b}+\Omega_\nu$ is kept
  constant. $N_\mathrm{CDM}$, $N_\mathrm{b}$ and $N_\nu$ correspond to
  the number of CDM, baryon and neutrino particles, respectively.
  All the simulations except P0 and P6 are based on the Fourier space
  implementation of neutrinos (see text).}
\label{tab_sims}
\end{table}

Our mock quasar spectra are based on
cosmological simulations run with the TreePM-SPH
code GADGET-3 \citep{Gadget2}. The code has been extended
to include neutrinos either by solving their potential on the mesh or
by representing them as discrete particles \citep{Springel}. Here, we use
primarily the first implementation and refer the reader to
\citep{Viel, Brandbyge_particle,Brandbyge_grid} for a critical comparison of
the two methods and also for comparison with another new method \cite{Bird_new}. Our 
simulations consist of $2\times512^3$ CDM plus
gas particles sampling a periodic box of $512~h^{-1}$Mpc. We adopt a
flat $\Lambda$CDM background with cosmological parameters
$\Omega_{\rm CDM}+\Omega_\nu=0.25$, $\Omega_\Lambda=0.7$, $\Omega_b=0.05$,
$h=0.7$ and $n_{\rm s}=1$. We consider three degenerates neutrino species
with a total mass of $\Sigma_i m_{\nu_i}=0.0, 0.3$ and $0.6$ eV.  The initial power 
spectra of most of our simulations, produced with 
CAMB\footnote{\url{http://camb.info/}}, are 
normalized for all neutrino masses at a wavenumber
$2\times 10^{-3} h~{\rm Mpc^{-1}}$, corresponding to the scale
constrained by CMB data. This produces different values of
$\sigma_8=0.877,~0.806$ and $0.732$ at $z=0$ for the models with
$\Sigma_i m_{\nu_i}=0.0,~0.3$ and $0.6$~eV, respectively (see Fig. 
\ref{power_spectrum}). Our initial conditions are generated at $z=49$. A summary of the 
different simulations we carried out is shown on table \ref{tab_sims}.

\begin{figure*}
\centering
\includegraphics[width=0.7\textwidth]{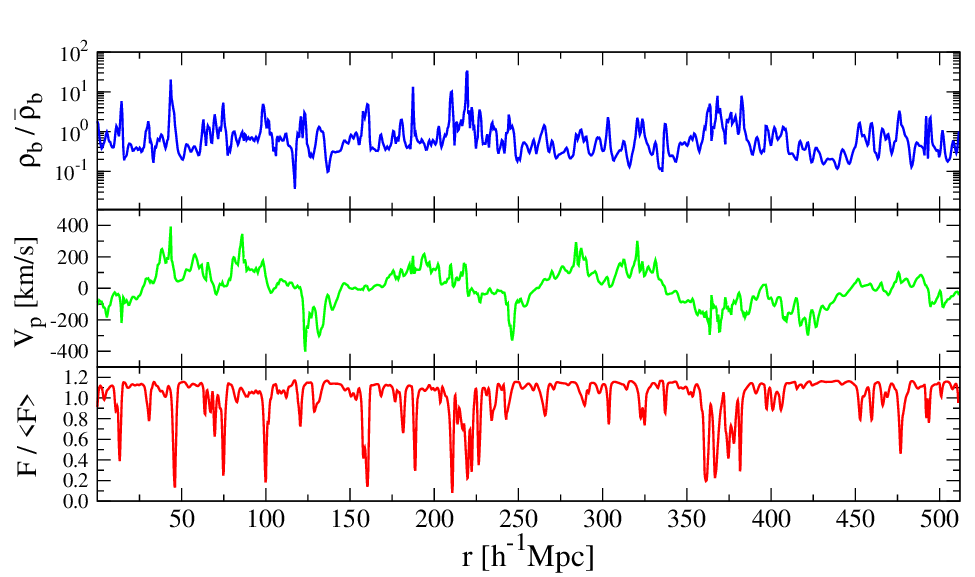}
\caption{Real space distribution of the baryon density contrast,  $\rho_b/\overline{\rho}_b$ (top panel), and the peculiar velocity,  $V_p$ (middle), along a random line-of-sight (RLOS). In the bottom panel we plot the transmitted flux $F=e^{-\tau}$, in units of the mean flux $\langle F \rangle$, in redshift space.}
\label{rho_vel_flux}
\end{figure*}

For each simulation we consider snapshots at redshifts $z=2.2$ and
$z=4$ that bracket the range of interest for the observed \lya~forest in
quasar spectra from ground-based telescopes.  For each snapshot we sample 4500 random line-of-sights (RLOSs) uniformly distributed along each $x,y$ or $z$ direction.  For each RLOS we extract the baryon
density contrast $\rho_b(r)/\overline{\rho}_b$ and the peculiar
velocity $V_p(r)$ along the line-of-sight and then compute the transmitted flux $e^{-\tau(u)}$ in redshift space (with $u$ in ${\rm km~s^{-1}}$), where $\tau$ is the \lya~optical depth, by using the
\textit{Fluctuating Gunn Peterson Approximation}:
\begin{equation}
\tau(u)=A\int_{-\infty}^{+\infty}dx~\delta[u-x-V_p(x)]~\bigg(\frac{\rho_b(x)}{\overline{\rho}_b}\bigg)^{1.6},
\end{equation}
where $x=H(z)r/(1+z)$ is the redshift space coordinate and $A$ is a factor that depends on the global thermal history of the IGM \citep{Croft},
\begin{eqnarray}
A=0.433\bigg(\frac{1+z}{3.5}\bigg)^6\bigg(\frac{\Omega_bh^2}{0.02}\bigg)^2\bigg(\frac{0.65}{h}\bigg)\bigg(\frac{3.68H_0}{H(z)}\bigg)\times\nonumber\\
\bigg(\frac{1.5\times10^{-12}s^{-1}}{\Gamma_{HI}}\bigg)\bigg(\frac{6000K}{T_0}\bigg)^{0.7},
\end{eqnarray}
with $\Gamma_{HI}$ being the hydrogen photoionization rate. The
power-law index in the scaling with $\rho_b/\overline{\rho}_b$ arises
from the equation of state for the IGM temperature,
$T=T_0(\rho_b/\overline{\rho}_b)^\alpha$ \citep{Hui}, with $\alpha
\approx 0.6$.  In all our calculations we adopt $T_0=10^4$ K and choose
$\Gamma_{HI}$ such that the mean flux over the whole set of RLOS
reproduce the observed mean flux at redshift $z$ \citep{Jordi}
$\langle F \rangle=e^{-\tau_{eff}(z)}$ with
$\tau_{eff}(z)=0.0023(1+z)^{3.65}$ \citep{Kim}. We neglect the effects
of thermal broadening. Finally, we smooth the flux over a scale of
$1~h^{-1}$Mpc, which is larger than the Jeans length, to avoid
sensitivity to substructure below the Jeans scale, which is affected by
numerical resolution and astrophysical processes (e.g. feedback from
galactic winds).

Figure \ref{rho_vel_flux} shows the baryon density contrast,
$\rho_b/\overline{\rho}_b$, and peculiar velocity, $V_p$, extracted
along a RLOS as a function of the comoving coordinate $r$ together
with the corresponding transmitted flux $F =e^{-\tau}$ in redshift
space, plotted in terms of the mean flux at redshift $z$.

\section{Analysis of the simulations} 

We focus our analysis on the statistical properties of low density
regions that are expected to produce weak absorption features.  We
define as void ``region'' a continuous domain in the transmitted flux
profile which remains always above a given threshold. The higher the
threshold, the lower the absorption in that region. For each RLOS we
extract the transmitted flux from $\rho_b/\mean{\rho}_b$ and $V_p$ and
count the number of regions above the selected threshold. This results
in a statistical estimate of the low absorption contribution to the
\lya$ $ signal, and allows us to quantify the impact of neutrinos on
those regions.

\begin{figure}
\centering
\includegraphics[width=0.49\textwidth]{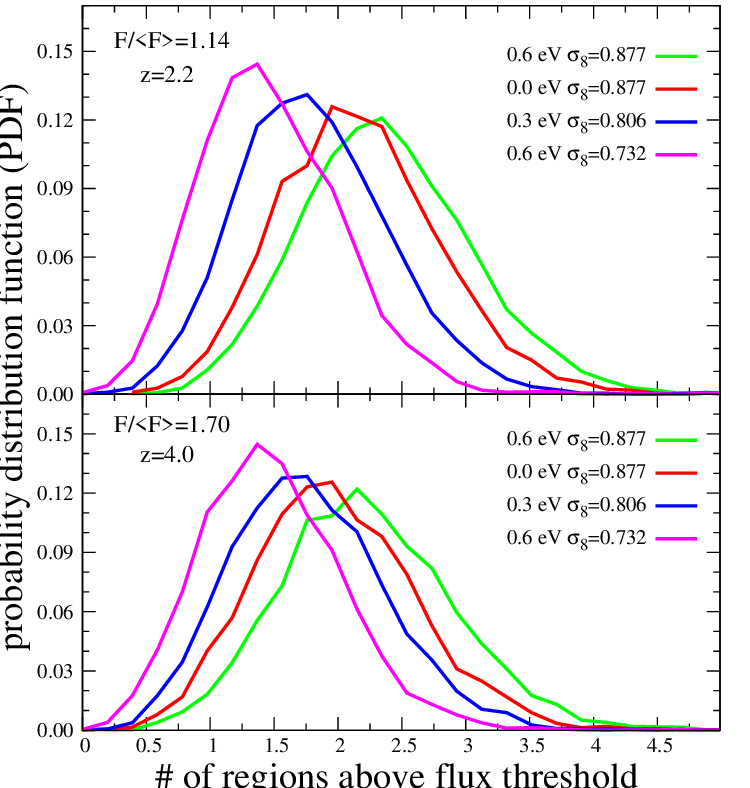}
\caption{Probability distribution function (PDF) for the number of    regions per path length of $100~h^{-1}$Mpc above a threshold of $F/\langle F\rangle=1.14$ (top), $1.70$ (bottom) as a function of  $\Sigma_i m_{\nu_i}$ and $\sigma_8$ at $z=2.2$ (top) and $z=4$ (bottom). The PDFs have long tails with a very low probability that extend up to 10-12. The $\sigma_8-\Omega_\nu$ degeneracy is not perfect and can be broken by studying the spectra at different redshifts.}
\label{above_pdf}
\end{figure}

\begin{figure}
\centering
\includegraphics[width=0.49\textwidth]{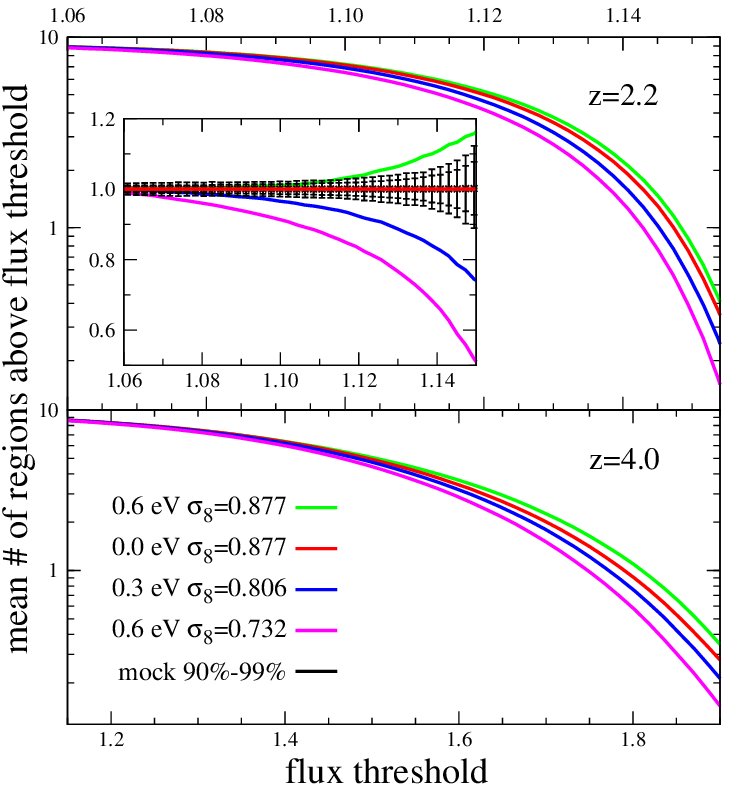}
\caption{Average number of regions per path length of $100~h^{-1}$Mpc as a function of flux threshold at
  redshift $z=2.2$ (top) and $z=4$ (bottom) for different
  neutrino masses and $\sigma_8$. The subplot in the upper panel shows the ratio between models with $\Sigma_i m_{\nu_i}\neq0.0$ and the model with $\Sigma_i m_{\nu_i}=0.0$. 
  The black error bars indicate the $90\%$ (interior tick marks) and $99\%$ (exterior tick marks) confident intervals for a mock catalog consisting of $200$ RLOS taken from the 
  simulation with $(\Sigma_i m_{\nu_i}=0.0$ eV, $\sigma_8=0.877)$. Models with $\Sigma_i m_{\nu_i}=0.3, 0.6$ and $\sigma_8=0.806, 0.732$ respectively can be ruled out with a high
  significance by using a catalog of 200 QSO spectra.}
\label{mean}
\end{figure}

In Fig. \ref{above_pdf} we plot the probability distribution function
(PDF) for the number of regions per path length of
$100~h^{-1}$Mpc \footnote{Non-integer numbers are due to the path
  length normalization.}  above a threshold of $F/\langle F\rangle
=1.14$ at redshift $z=2.2$ (top) and at redshift $z=4.0$ for a
threshold $F/\langle F\rangle =1.70$ (bottom) for three different
neutrino masses, $\Sigma_i m_{\nu_i}=0.0,~0.3,~0.6$ eV. The two numbers above are 
chosen taken into account two competing effects: on one side, the larger the value of $F/
\langle F\rangle$ the larger the differences between the models. On the other side, the 
number of regions above the threshold drops rapidly as $F/\langle F\rangle$ increases, 
requiring a larger QSO spectra catalog to obtain converged results. By choosing the 
numbers above we make sure that differences are large enough having converged 
results. We have
verified that these PDFs do not change if we increase the number of
RLOS, i.e. our statistical sample of RLOS is large enough to reliably
measure the PDF. Figure \ref{above_pdf} shows that the neutrino mass
has a significant impact on the mean of the distributions.  In
Fig.~\ref{mean} we plot the mean of the distributions, i.e. the
average number of regions per path length of $100~h^{-1}$Mpc above a
given threshold, as a function of the threshold for the three
different neutrino masses ($\Sigma_i m_{\nu_i}=0.0,~0.3,~0.6$ eV) at
redshift $z=2.2$ (top) and $z=4.0$ (bottom).  This shows clearly that
the higher the threshold, the larger the differences between the
various neutrino cosmologies. This is the expected neutrino signature
as we discuss below.


We explicitly checked that relative differences between our neutrino
models are numerically converged against mass and spatial resolution.
Furthermore, we used the neutrino particle implementation (simulations P0 and P6 on 
table \ref{tab_sims}) and found
the same trends in the neutrino signature as with the grid method,
although relative differences between the different models are even
slightly larger when we use the particle implementation.  This is due
to the fact that non-linear neutrino effects, such as phase mixing,
are only properly captured by using the particle implementation. However, we
note that the grid implementation in the mildly non-linear \lya~
regime is fully justified since non-linear neutrinos effects should
not be particularly important at those redshifts and at $k<1~h~{\rm
  Mpc^{-1}}$.

 
\section{Systematic errors} 
 
The high transmissivity regions of the Lyman-$\alpha$ forest are prone
to systematic errors such as those induced by the continuum fitting
procedure and the signal to noise ratio ($S/N$) of the QSO spectrum. We have
investigated whether these effects are able to spoil the neutrino
signal. In particular, we have studied how the presence of systematic errors affects
the differences between models (in terms of the average number of
regions above the threshold at $z=2.2$), and their impact on the
 sensitivity to $\Omega_\nu$ of a catalog containing 200 QSO spectra.

We first focus our attention to the case without systematic errors. In
the subplot of Fig. \ref{mean} we show the average number of regions
per path length of $100~h^{-1}$Mpc as a function of the threshold at
$z=2.2$ normalized to the neutrinoless model. The black error bars
show the $90\%$ (interior tick marks) and $99\%$ (exterior tick marks)
confident intervals for a mock catalog consisting of $200$ RLOS taken
from the simulation with $(\Sigma_i m_{\nu_i}=0.0$ eV,
$\sigma_8=0.877)$. We find that with a catalog consisting of $200$ QSO
spectra (not affected by systematic errors) we can rule out models
$(\Sigma_i m_{\nu_i}=0.3$ eV, $\sigma_8=0.806)$ and $(\Sigma_i
m_{\nu_i}=0.6$ eV, $\sigma_8=0.732)$ with a high
significance. Distinguishing models with the same $\sigma_8$ would
require a larger QSO spectra catalog and combining results at
different redshifts in order to disentangle the different redshift
evolution of the models.

In order to study how systematic errors impact on our results, we have
created a realistic mock catalog of high resolution QSO spectra. We
have mimicked the continuum fitting errors by rescaling each flux
pixel by the quantity $F_i/F_{max}$ ($F_{max}$ being the maximum value
of the transmitted flux along the spectrum) as done by
\citep{McDonald}; we considered a $S/N=100$ which is reasonable for
UVES/VLT QSO spectra. Note that this treatment of the systematic
errors induced by the continuum fitting is rather simplistic and
conservative, and ideally one would like to put more refined models
for the QSO continuum and generate a set of mock spectra that would be
as close as possible to the observed one (for example by including the
redshift distribution of the sources), however this is beyond the
scope of the present paper.

\begin{figure}
\begin{center}
\includegraphics[width=0.49\textwidth]{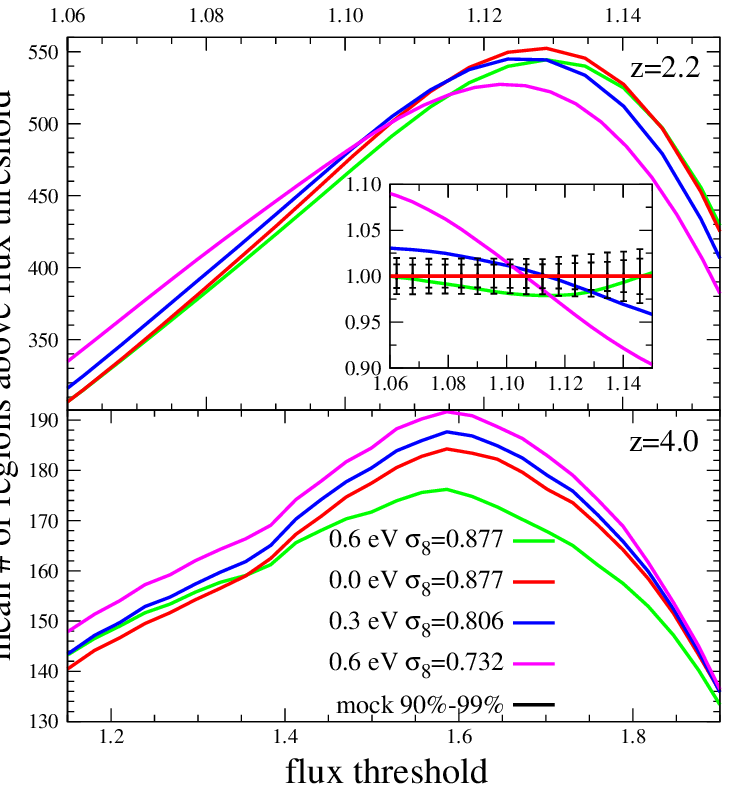}
\end{center}
\caption{Effects of the Gaussian noise in the QSO spectra and the
  continuum fitting procedure on the properties of the high
  transmission regions of the \lya~ forest. We create high resolution
  ($S/N=100$) mock QSO spectra mimicking
  the continuum fitting errors by using the prescription of McDonald
  et al. (2000). In the figure we plot the average number of regions
  per path length of 100 $h^{-1}$Mpc as a function of threshold at
  redshifts $z=2.2$ (top) and $z=4.0$ (bottom) for different neutrino
  masses and $\sigma_8$. The subplot shows the ratio between models
  with $\Sigma_i m_{\nu_i}\neq0.0$ and the model with $\Sigma_i
  m_{\nu_i}=0.0$ at $z=2.2$. The black error bars indicate the $90\%$
  (interior tick marks) and $99\%$ (exterior tick marks) confident
  intervals for a mock catalog consisting of $200$ high resolution QSO
  spectra taken from the simulation with $(\Sigma_i m_{\nu_i}=0.0$ eV,
  $\sigma_8=0.877)$. Systematic errors do not erase the neutrino
  signal, however, the differences between models become smaller. At
  $z=2.2$, only the model with $\Sigma_i m_{\nu_i}=0.6$ eV and
  $\sigma_8=0.732$ can be ruled out with a high significance using the
  catalog considered in this example.}
\label{mean_above_noise}
\end{figure}

Using this
catalog, we have repeated the analysis described above and we show the
results in Fig. \ref{mean_above_noise}. Two things can be pointed out
from this figure. On one side, we find that the mean number of regions
above a flux threshold is strongly affected by the Gaussian noise and
by the bin size in the QSO spectra. This dependence can be easily
understood considering that the Gaussian noise on a pixel can divide a
single region above a threshold into two, and also by the fact that
spurious regions will appear because the Gaussian noise can increase
the value of the transmitted flux (ending up with a final value above
the threshold) on a pixel which is below the threshold. On the other
side, it turns out that, as expected, the differences between models
become smaller as the QSO catalog $S/N$ drops. However, the
remaining differences between models points out that neutrino effects
are not erased by the presence of the systematics discussed in this
Section. The subplot of Fig. \ref{mean_above_noise} shows the ratio
between the different models and the model with $\Sigma_i
m_{\nu_i}=0.0$. The black error bars indicate the $90\%$ (interior
tick marks) and $99\%$ (exterior tick marks) confident intervals for a
mock catalog consisting of $200$ high resolution QSO spectra extracted
from the simulation with $(\Sigma_i m_{\nu_i}=0.0$ eV,
$\sigma_8=0.877)$. We find that with a high resolution catalog
($S/N=100$) consisting of 200 QSO spectra, models with $\Sigma_i
m_{\nu_i}=0.0$ and $\sigma_8=0.877$ and $\Sigma_i m_{\nu_i}=0.6$ and
$\sigma_8=0.732$ can be distinguished at a very high significance. The
models with $\Sigma_i m_{\nu_i}=0.0$ and $\sigma_8=0.877$ and
$\Sigma_i m_{\nu_i}=0.3$ and $\sigma_8=0.806$ can be distinguished
with a lower significance, while in order to distinguish models with
the same $\sigma_8$, a catalog with either more QSO spectra or higher
$S/N$ value is needed.

In addition to the systematic errors arising from the noise in the
spectra and the continuum fitting procedure, the presence of metal
lines should not affect strongly our findings since in high resolution
spectra these lines can be identified and metal-free regions can be
conservatively used in the analysis and, by smoothing the transmitted
flux over a region which is typically $\sim 1$ com. $h^{-1}$Mpc
(roughly twenty times larger than the typical width of a metal line),
we should be less sensitive to these contaminants.

\section{Sensitivity to neutrino masses using the Fisher Matrix formalism} 

In this section we quantify the sensitivity to $\Omega_\nu$ of the
observable described in this paper. 

From Figs. \ref{above_pdf}, \ref{mean} and \ref{mean_above_noise} it
is clear that both $\Omega_\nu$ and $\sigma_8$ impact on the
statistics of the high transmission regions of the Lyman-$\alpha$
forest. Variations in the mean flux, in the thermal and ionization
history of the IGM are also expected to impact on the Lyman-$\alpha$
properties of large size voids \citep{Viel_voids}. For real data, the
signal to noise ratio is not known with infinite precision (a 10\%
error could be a reasonable conservative assumption), and as can be
seen when comparing Figs. \ref{mean} and \ref{mean_above_noise},
variations in it are expected to strongly impact on the statistics of
low density regions. The scale over which we smooth the transmitted
flux spectra will also impact on our results.  However, we have full
control on this scale, that we can set to a any particular value.
Therefore we do not need to consider this parameter in this analysis.

We use the Fisher matrix formalism to study how the above parameters
affect the statistics of the high transmissivity regions of the
Lyman-$\alpha$ forest and with which error can $\Omega_\nu$ be
constrained by using the method proposed in this paper.  The
parameters, $\vec{p}=(p_1,p_2,p_3,p_4)$, we use in the analysis are:
the sum of the neutrino masses, $\Sigma_i m_{\nu_i}$, the catalog mean
transmitted flux, $\langle F \rangle$, the parameter $\alpha$, present
in the IGM temperature-density relation,
$T=T_0(\rho_b/\overline{\rho}_b)^{\alpha}$, and $\sigma_8$. Note that
the parameters $T_0$ and $\Gamma_{HI}$ affect the properties of the
Lyman-$\alpha$ forest only through $\langle F \rangle$, and therefore,
we do not need to include them in the analysis. For the realistic
catalog, the one that incorporates the systematic errors, we also need
to consider a further parameter, $p_5$, which corresponds to the
catalog $S/N$.

We carry out the Fisher matrix analysis for two different catalogs. In
the first one we consider a catalog consisting in 200 QSO spectra with
no continuum fitting errors and with a $S/N$ equal to infinite. This
catalog, although unrealistic, help us to determine the tightest
constrains on $\Omega_\nu$ we can achieve by using the method here
proposed. For the second one, we use a 200 QSO spectra catalog with
$S/N$=100. The continuum fitting
errors are mimicked using the prescription of \cite{McDonald} (see
section 4). In both catalogs, we smooth the transmitted flux spectra
over a scale of 1 $h^{-1}$ com. Mpc, which is comparable to the Jeans
scale. This smoothing is intended in order to be less sensitive to the
substructure below this scale and to the noise level properties and
can be applied on both simulations and real spectra in exactly the
same way.

Our observables, $f_b$, with b=0,1,2,....,29, correspond to the
average number of regions per path length of 100 $h^{-1}$ Mpc above a
threshold equal to $F_b=0.90+0.09b/29$. We assume flat priors on the
parameters, and therefore, the posterior distribution is equal to the
likelihood. The likelihood associated to each model will be given by

\begin{equation}
\mathcal{L}\propto \rm{exp}\big[-\frac{1}{2} \displaystyle\sum_{a=0}^{29} \displaystyle\sum_{b=0}^{29} (f_a-\widetilde{f}_a(\vec{p}))C^{-1}_{ab} (f_b-\widetilde{f}_b(\vec{p}))\big]~,
\end{equation}

where $\widetilde{f}_b(\vec{p})$ is the theoretical prediction for $f_b$ for a model with parameters $p_i$; $C$ is the covariance matrix

\begin{equation}
C_{ab}=\langle(f_a-\widetilde{f}_a) (f_b-\widetilde{f}_b)\rangle~.
\end{equation}


Let $\vec{p}^0$ be the values that maximize the likelihood. Around $\vec{p}^0$ we can expand $\rm{ln}~\mathcal{L}$ in a Taylor series as

\begin{equation}
\rm{ln}~\mathcal{L}=\rm{ln}~\mathcal{L}(\vec{p}^0)+\frac{1}{2}\displaystyle\sum_{ij}(p_i-p_i^0)\frac{\partial^2\rm{ln}~\mathcal{L}}{\partial p_i \partial p_j}(p_j-p_j^0)+...~
\end{equation}

Note that because the likelihood has a maximum in $\vec{p}^0$, $\partial \rm{ln}~\mathcal{L}/\partial p_i(\vec{p}^0)=0$. The Fisher matrix is defined as

\begin{equation}
F_{ij}=\left \langle \frac{\partial^2\rm{ln}~\mathcal{L}}{\partial p_i \partial p_j}\right \rangle~,
\end{equation}

and the marginalized error in the parameter $p_i$ satisfies the relation

\begin{equation}
\sigma_{p_i}\geqslant\sqrt{F_{ii}^{-1}}~.
\end{equation}

If the data are distributed according to a Gaussian distribution, the Fisher matrix can be written in the following way \citep{Tegmark,Heavens}

\begin{eqnarray}
F_{ij}=\frac{1}{2}\rm{Tr}[C^{-1}\partial_iCC^{-1}\partial_jC&+&C^{-1}\partial_i \widetilde{f}_b(\vec{p}) \partial_j (\widetilde{f}_b(\vec{p}))^T\\
&+&C^{-1}\partial_j \widetilde{f}_b(\vec{p})\partial_i (\widetilde{f}_b(\vec{p}))^T]
\end{eqnarray}

\begin{equation}
F_{ij}=\frac{1}{2}\rm{Tr}[C^{-1}\partial_iCC^{-1}\partial_jC]+\displaystyle\sum_{a=0}^{29} \displaystyle\sum_{b=0}^{29} \frac{\partial\widetilde{f}_a}{\partial_{p_i}}C^{-1}_{ab}\frac{\partial\widetilde{f}_b}{\partial_{p_j}}
\label{fisher_eq}
\end{equation}

where $\partial_i=\partial/\partial p_i$ and Tr is the matrix trace. We find that the first term 
in the previous equation contributes very little to the Fisher matrix. This happens because 
the errors (the covariance matrix) depend very weakly of the parameters $\vec{p}$. For 
that reason, we have neglected that term in our analysis\footnote{We have explicitly checked that the presence of this term do not change our results}. 

\begin{figure*}
\begin{center}
\includegraphics[width=1.0\textwidth]{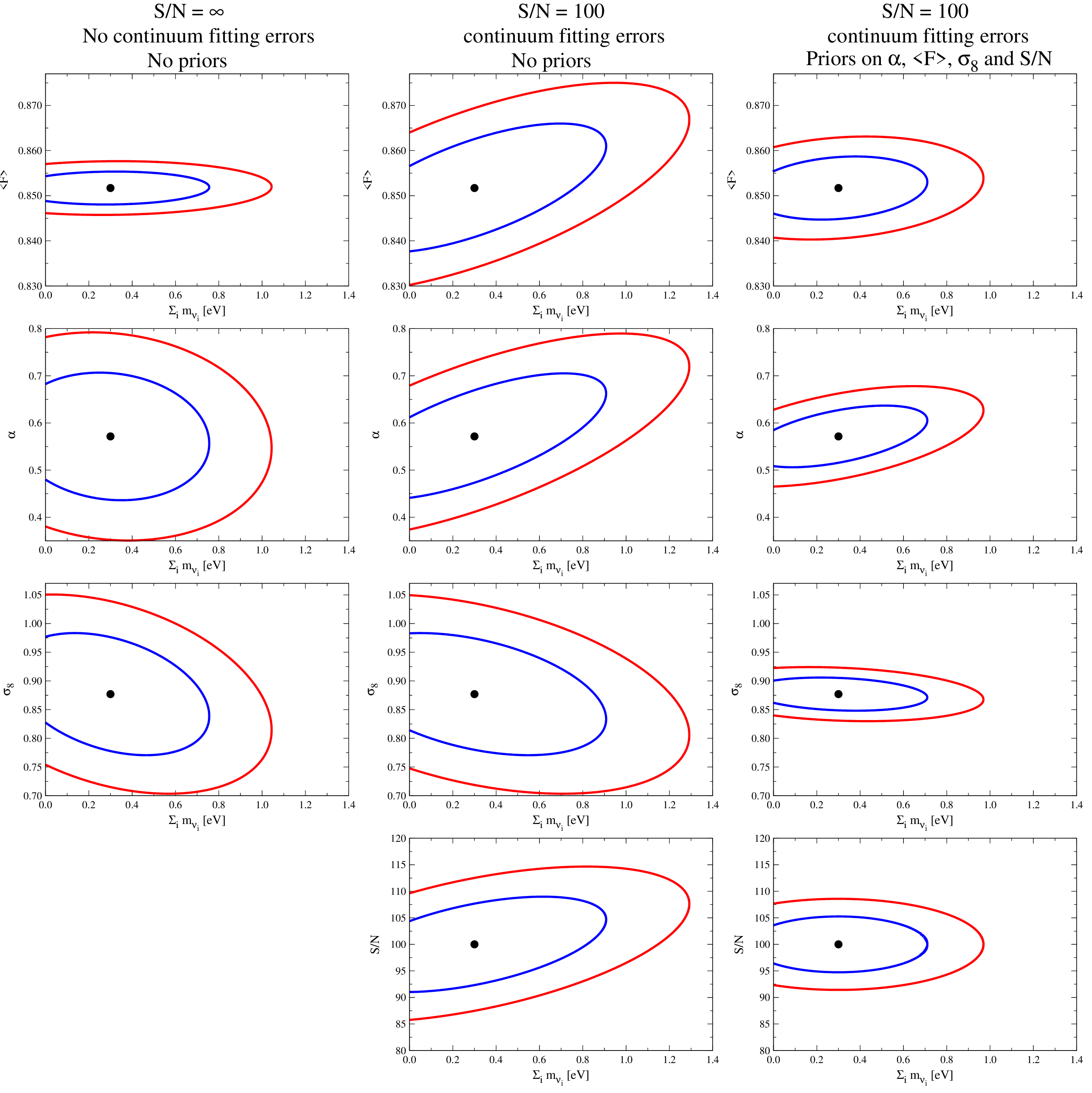}
\caption{Contour plots at $1\sigma$ (blue) and $2\sigma$ (red) showing the correlation between $\Sigma_i m_{\nu_i}$ and $\alpha$, $\langle F \rangle$, $\sigma_8$ and $S/N$ for a catalog consisting in 200 QSO spectra for three different situations: a catalog with $S/N=\infty$ and without continuum fitting errors, assuming no priors on the parameters (left column), a catalog with $S/N=100$ and without assuming priors on the parameters (middle column) and a catalog with $S/N=100$  assuming priors of $10\%$ in the value of $\alpha$ and $3\%$ in the value of $\sigma_8$ (right column). The black points show the position of the fiducial model.} 
\end{center}
\label{FoM}
\end{figure*}

The values of the parameters for our fiducial model are:
$p_1^0=\Sigma_i m_{\nu_i}=0.3$ eV, $p_2^0=\langle F \rangle=0.8517$,
$p_3^0=\alpha=0.5714$ and $p_4^0=\sigma_8=0.877$. For the realistic catalog, the fifth parameter has a fiducial value equal to $p_5^0=S/N=100$.
We first computed the Fisher matrix for a catalog
with a $S/N=\infty$, i.e. for the ideal case, unaffected by systematic
errors, that we consider in the body of this article. Without using
priors on the parameters, we find that the marginalized error in
$\Sigma_i m_{\nu_i}$ is roughly given $0.30\sqrt{200/N}$ eV, where $N$
is the number of QSO spectra in the catalog. In
the left column of Fig. \ref{FoM} we show the contour plots at $1\sigma$
(blue) and $2\sigma$ (red) for a catalog consisting of 200 QSO
spectra with $S/N=\infty$.

We have also studied the sensitivity of our method making one further assumption: we suppose that the amplitude of the matter power spectra is fixed on large scales. By making that assumption, $\Omega_\nu$ and $\sigma_8$ are no longer independent parameters. We have carried out the Fisher matrix analysis taking as parameters $\vec{p}=(\Sigma_i m_{\nu_i},\langle F \rangle, \alpha)$ and we found that the marginalized error in $\Sigma_i m_{\nu_i}$ is $\sim0.25\sqrt{200/N}$ eV. Although we have one parameter less than in the previous case, the marginalized error in $\Omega_\nu$ is not significantly reduced. This happens because we find a strong
correlation between the parameters $\Sigma_i m_{\nu_i}$ and
$\alpha$. This correlation is the main source of error when computing
the marginalized error in $\Sigma_i m_{\nu_i}$.  We note that this
correlation is expected and has a physical meaning: a larger value of
$\alpha$ makes the low density IGM colder and thereby would result in
a larger amount of neutral hydrogen. In such a case the voids will
contain more matter and this is the same effect that can be achieved
in a universe with a larger value of $\Sigma_i m_{\nu_i}$. We note that this huge 
degeneracy between $\Sigma_i m_{\nu_i}$ and $\alpha$ is broken once we introduce $
\sigma_8$ in the Fisher matrix. This can be understood by looking at Fig. \ref{mean}: the 
mean number of regions above the threshold varies differently depending on whether $
\sigma_8$ is fixed (compare red and green lines) or the amplitude of the power spectra is 
fixed on large scales (compare red, blue and magenta lines).

By repeating the Fisher matrix analysis with the 200 high resolution QSO spectra catalog (S/N$\sim$100), without assuming priors
on the parameters, we find that the marginalized error in $\Sigma_i
m_{\nu_i}$ is given by $0.4\sqrt{200/N}$ eV. In the middle column of
Fig. \ref{FoM} we show the contour plots for a catalog consisting in
200 high resolution QSO spectra. 

In order to investigate the tightest constraints on $\Sigma_i m_{\nu_i}$
that we can achieve with this catalog, we impose realistic priors on the
parameters. We assume that $\langle F \rangle$, $\alpha$ and the $S/N$
are known within a $10\%$ error and that $\sigma_8$ is known within
$3\%$. The previous error intervals are at $1\sigma$. We repeat the
Fisher matrix analysis and we find a marginalized $1\sigma$ error on
$\Sigma_i m_{\nu_i}$ equal to 0.27 eV for a catalog consisting in 200
high resolution QSO spectra. The right column of Fig. \ref{FoM} show
the contour plots between $\Sigma_i m_{\nu_i}$ and the other
parameters.

Finally, we repeat the whole analysis for a different fiducial model with
$p_1^0=\Sigma_i m_{\nu_i}=0.0$ eV (the values of the other parameters
are the same as those of the above fiducial model) and we find that
the marginalized error in $\Omega_\nu$ and the correlation with the
other parameters are in very well agreement with those obtained for
the 0.3 eV above fiducial model. This reinforces our assessment that
the contribution of the first term in Eq. \ref{fisher_eq} to the whole
Fisher matrix is negligible.
 
\section{Discussion and Conclusions} 

We have studied in detail the sensitivity of the threshold crossing
statistics (see for example \citep{Jordi, Fan}) to the masses of the
neutrinos. We have focused our study on the low density regions of
the Lyman-$\alpha$ forest in quasar spectra. Those regions correspond
to the innermost parts of non-linear voids. We find that the number of
regions above a given threshold in the flux is strongly affected by
the masses of the neutrinos. The changes between different models are due to two
factors: the change in amplitude and slope in the linear power
spectrum driven by neutrinos and non-linear effects associated with
CDM and neutrinos (note that neutrinos modify the non-linear evolution
of the CDM distribution). The inner panel of Fig. \ref{power_spectrum}
shows the linear (solid lines) and non-linear (dotted lines) versions
of the power spectrum at $z=3$ normalized to the case without
neutrinos. Whereas the modification on large scales ($k<0.03~h~{\rm
  Mpc^{-1}}$) is due to the change in the linear power spectrum, we
find that on smaller spatial scales the non-linear effects
dominate. 

By creating realistic mock catalogs of high resolution QSO spectra, we
have found that systematic errors, such as the noise in the spectrum
of the continuum fitting procedure, reduce the differences between
different models (in terms of the average number of regions above the
threshold as a function of the threshold) but do not erase the
neutrino signal in the high transitivity regions of the Lyman-$\alpha$
forest. We have used the Fisher matrix formalism to forecast the
errors associated to a measurement of $\Omega_\nu$ using the method
proposed in this paper. We conclude that $\Sigma_i m_{\nu_i}$ can be
measured with an error of 0.4 eV ($1\sigma$) using a catalog of 200
high resolution QSO spectra. By assuming that the values of the
parameters $\langle F \rangle$, $\alpha$ and the $S/N$ are known with
a $10\%$ error and that the value of $\sigma_8$ is known with a $3\%$
error, we find that the $\Sigma_i m_{\nu_i}$ can be measured with a
precision of 0.27 eV ($1\sigma$) by using 200 high resolution QSO
spectra.  This is not an unreasonable number of high resolution
spectra and is already available to the community.  This accuracy is
achievable given the current efforts to measure the IGM thermal state
by using the flux PDF and power spectrum, wavelets and the line-width
distribution from Voigt profile fitting.

We note that the statistics we have studied here to measure the
neutrino masses implicitly contains more information than the one that
can be extracted from measurements of the power spectrum. The
threshold crossing statistics is analogous to the genus curve used to
characterize the topology of the three-dimensional galaxy distribution
\citep{Fan}. The amplitude of the genus per unit of volume depends
only on the second moment of the power spectrum  \citep{Hamilton}
if the field is Gaussian. For non-Gaussian fields (as the ones we are
studying here) the amplitude of the genus curve contains information
of higher order correlation functions (see for example \citep{Zhang}).

The aim of this method is to put a new and independent constrain on
$\Omega_\nu$ using the Lyman-$\alpha$ forest. The results found with
this method can be combined with other cosmological measurements such
as the CMB or LSS to improve current bounds on the masses of the
neutrinos. Previous studies \citep{Viel-Enzo} have demonstrated the
advantages of using the flux PDF or void statistics rather than the
power spectrum or bispectrum to distinguish cosmological models with
small differences (in this case authors investigated
non-Gaussianities).

We have also studied another set of statistics which carry more
information than the one compressed in the matter power spectrum. One
of these statistics\footnote{They are called \textit{threshold
    probability functions.}}, $C_2(R,F_{th})$, is widely used in
material science \citep{Torquato, Jiao}, and has been recently applied
to the Lyman-$\alpha$ forest \citep{KGLee}. For the Lyman-$\alpha$
forest, $C_2(R,F_{th})$, known as the cluster function, is defined as
the probability function of finding a pair of pixels in the same
phase, belonging to the same region, separated by a distance $R$. The
pixels in the transmitted flux spectrum are assigned to two different
phases: phase 1 if the value of the transmitted flux is larger than
$F_{th}$ and phase 2 otherwise. We have studied a statistics directly
related to $C_2(R,F_{th})$: the probability distribution function of
the sizes of the connected regions above a given threshold in the
transmitted flux spectrum. We have also focused on the high
transitivity regions of the Lyman-$\alpha$ forest, restricting our
study to the phase 1, to $z=2.2$ and to values of $F_{th}$ larger than
0.9. We have found that neutrino masses leaves an imprint in this
statistics, although it is smaller than the one we have presented in
this paper. By using the Fisher matrix formalism, we conclude that
this method could distinguish neutrino masses with a precision about
0.4 eV using 200 QSO spectra unaffected by systematic errors. This
value should be compared with the error equal to 0.3 eV that we
obtained with our method for the same number of QSO spectra. For this
reason, we believe that the statistics we have presented in this paper
is one of the most suitable observables, containing higher order
information, to place new independent bounds on the masses of the
neutrinos, even if the $1\sigma$ error bar is less competitive than
that obtained by other cosmological probes.

Given that the error in $\Sigma_i m_{\nu_i}$ can be significantly
reduced by: adding priors to the parameters, measuring independently
the IGM thermal history and/or using QSO spectra with $S/N$ larger
than 100, we conclude that in the near future, a large (but not
unreasonable) number of high resolution QSO spectra could provide a
relatively tight, new and independent constraint on neutrino masses
which will be complementary to that provided by other large scale
structure probes.

\noindent

\section*{Acknowledgments} 
We thank the referee for the very helpful report.  We thank Enzo
Branchini, Olga Mena, Jordi Miralda-Escud\'e, Carlos Pe\~na-Garay,
Khee-Gan Lee, Jonathan Pritchard and Jes\'us Zavala for discussions
and helpful comments on the manuscript and Volker Springel for
providing us with GADGET-3.  The simulations were run on the Darwin
Supercomputer center HPCF in Cambridge (UK) and at the research
computing center of Harvard University. FVN was supported by the CSIC
under the JAE-predoc program when this work started. This work was
supported in part by grants: ASI/AAE, INFN/PD51, ASI/EUCLID, PRIN-INAF
2009, PRIN-MIUR (for MV), as well as NSF grant AST-0907890 and NASA
grants NNX08AL43G and NNA09DB30A (for AL). MV and FVN are supported by
the ERC Starting Grant ``CosmoIGM''.

\bibliographystyle{mn2e} \bibliography{master2.bib}

\end{document}